\documentclass[12pt]{article}

\usepackage{amssymb}

\begin{document}
\def\b{\beta}\def\d{\delta}\def\g{\gamma}
\def\a{\alpha}\def\t{\tau}\def\l{\lambda}
\def\e{\epsilon}\def\r{\rho}\def\d{\delta}
\def\wt{\widetilde}\def\wid{\widehat}
\def\ds{\displaystyle}\def\ov{\overline}
\def\be{\begin{equation}}\def\ee{\end{equation}}
\def\beq{\begin{eqnarray}}\def\eeq{\end{eqnarray}}
\def\om{\omega}\def\p{\partial}\def\s{\sigma}
\def\k{\varkappa}\def\mf{\mathcal{F}}
\makeatletter
\@addtoreset{equation}{section}
\makeatother
\renewcommand{\theequation}{\thesection.\arabic{equation}}
\newtheorem{thm}{Theorem}[section]
\newtheorem{prop}[thm]{Proposition}
\newtheorem{lem}[thm]{Lemma}
\newtheorem{fact}[thm]{Fact}
\newtheorem{conj}[thm]{Conjecture}

\title{Separation of variables for the $A_3$ elliptic
Calogero-Moser system}

\author{V.V. Mangazeev\thanks{E-mail: vladimir@maths.anu.edu.au}\\
\itshape Centre for Mathematics and its Applications, \\ \itshape
School of Mathematical Sciences, \\ \itshape
 The Australian National University,\\ \itshape
ACT0200, Australia}
\date{January 16, 2001}

\maketitle

\begin{abstract}
We consider the classical elliptic Calogero-Moser model. A set of
canonical separated variables for this model has been constructed
in \cite{KNS}. However, the generating function of the separating
canonical transform is known only for two- and three-particle
cases \cite{KNS}. We construct this generating function for the
next $A_3$ case as the limit of the conjectured form of the quantum
separating operator. We show explicitly that this generating
function gives a canonical transform from the set of original
variables to the separated ones.

\vspace{0.5cm}

\end{abstract}

\newpage

\section{Introduction}

The separation of variables method (SoV) is one of the powerful
approaches to solve spectral problems for quantum integrable
systems (see \cite{S} for an overview). This method was
successfully  applied to many integrable systems. However, it
appears that the Calogero-Moser system (CMS) \cite{C,M} (and its
relativistic analog: Ruijsenaars model \cite{R}) is the example
where SoV encounters some difficulties. Namely, as it was shown
in \cite{S1} the classical $r$-matrix for the CMS depends
explicitly on dynamical variables when a quantization procedure
is not known. As a result  all approaches concerning the quantum
separation of variables for these systems used, in fact, {\it ad
hoc} methods. The interest to produce a separation of variables
for  the CMS is two-fold. The first reason is obvious: the SoV
method can help to reduce the multi-dimensional spectral problem
for the CMS to a set of one-dimensional ones which are easier to
handle. The second reason is a connection of different limits of
the CMS with symmetric functions \cite{Ma}. In particular, the
separation of variables method for the $A_2$ quantum CMS in the
trigonometric limit produces a new integral representation for
the $A_2$ Jack polynomials \cite{KS}.

In the paper \cite{KNS} a set of separated canonical variables
has been constructed for the classical Ruijsenaars model (
which gives CMS when $\l\to0$). The new canonical
separated variables come as poles of the properly normalized
Baker-Akhiezer function. However, to describe explicitly a transformation
to the new set of canonical variables we need to know
a generating function. This function was constructed in \cite{KNS}
for the $A_2$ case. For the $A_n$, $n>3$ it satisfies
complicated nonlinear partial differential equations (PDE) and there
is a little hope to solve them directly. In this paper
we will show how to parameterize solutions of this PDE
for the $A_3$ case. This parameterization comes naturally from the
asymptotics of the solutions of special systems of linear
equations for the separating kernel in the quantum case.

The paper organized as follows. In Section 2 we remind the
main properties of the classical Calogero-Moser system
and give necessary definitions of Weierstrass elliptic functions.
In Section 3 we remind  the main facts about
the separation of variables \cite{KNS} for the CMS and introduce some
convenient notations. In Section 4 we formulate the quantum
version of the model and make a conjecture on the quantum
separation operator. In Section 5 we prove the main theorem
that the $A_3$ generating function is given by the asymptotics of
this operator. In Section 6 we give some concluding remarks.

\section{ The classical Calogero-Moser system}

The elliptic $N$-particle Calogero-Moser system \cite{C,M}
is described in
terms of canonical variables $p_i$, $q_i$, $i=1,\ldots,N$ with
Poisson brackets

\be
\{p_i,p_j\}=\{q_i,q_j\}=0,\quad \{p_i,q_j\}=\d_{ij} \ee and
Hamiltonian
\be
{\cal{H}}=\sum_{i=1}^Np_i^2+g^2\sum_{i\neq
j}\wp(q_i-q_j),\label{ham} \ee where $\wp(x)$ is the Weierstrass
function with periods $2\om_1$ and $2\om_2$ (see, for example,
\cite{BE}).

Let us summarize some important properties of Weierstrass
functions \cite{BE} to be used later. We define the Weierstrass
$\s$-function by the infinite product \be
\s(x)=x\prod_{m,n\neq0}(1-{x\over\om_{mn}})\exp\Biggl[{x\over\om_{mn}}
+{1\over2}\Bigl({x\over\om_{mn}}\Bigr)^2\Biggr] \ee where
$\om_{mn}=2m\om_1+2n\om_2$, $m,n\in {Z}$ and $\Gamma=2\om_1
Z+2\om_2 Z$ is the period lattice.
 Then $\zeta$ and $\wp$
Weierstrass functions are defined as
\be
\zeta(x)={\s'(x)\over\s(x)}    ,\quad \wp(x)=-\zeta'(x). \ee

The function $\wp(x)$ is an elliptic function of periods $2\om_1$,
$2\om_2$, which is even and has the only double pole at $z=0$
in the primitive domain
${\mathfrak{D}}:=\{z=2\om_1 x+2\om_2y|x,y\in[0,1)\}$.
The functions $\zeta(x)$ and $\s(x)$ are odd functions, which are
quasi-periodic, obeying
\be
\zeta(x+2\om_{1,2})=\zeta(x)+2\eta_{1,2},\quad \s(x+2\om_{1,2})=
-\s(x)e^{2\eta_{1,2}(x+\om_{1,2})},\label{period} \ee where
$\eta_{1,2}=\zeta(\om_{1,2})$ and
$\eta_1\om_2-\eta_2\om_1={i\pi\over2}$.

They have the following expansions near the origin \be
\s(x)=x-{g_2x^5\over240}-{g_3x^7\over840}-{g_2^2x^9\over161280}+...,\>\>
\zeta(x)={1\over
x}-{g_2x^3\over60}-{g_3x^5\over140}-{g_2^2x^7\over8400}+...\label{expa}
\ee with \be g_2=60\sum_{m,n\neq0}{1\over\om_{mn}^4},\quad
g_3=60\sum_{m,n\neq0}{1\over\om_{mn}^6}. \ee

Weierstrass functions satisfy  addition theorems, the most
important are \be
\zeta(x+y)=\zeta(x)+\zeta(y)+{1\over2}{\wp'(x)-\wp'(y)\over\wp(x)-\wp(y)},
\label{add1}\ee \be
\wp(x+y)+\wp(x)+\wp(y)=[\zeta(x+y)-\zeta(x)-\zeta(y)]^2
\label{add2}\ee \be
\s(x+y)\s(x-y)=-\s^2(x)\s^2(y)[\wp(x)-\wp(y)], \ee
\be\Phi(u,x)\Phi(u,y)=\Phi(u,x+y)[\zeta(u)+\zeta(x)+\zeta(y)-\zeta(u+x+y)]
\label{addth},\ee where \be
\Phi(u,x)={\s(u+x)\over\s(u)\s(x)}\label{phi} \ee and the
generalized Cauchy identity \be \mbox{\rm
det}\Bigl[\Phi(u,x_i-y_j)\Bigr]=\Phi(u,\Sigma)\s(u,\Sigma)
{\prod_{k<l}\s(x_k-x_l)\s(y_l-y_k)\over\prod_{k,l}\s(x_k-y_l)} \ee
with ${\ds\Sigma=\sum_i(x_i-y_i)}$.

 The system with hamiltonian (\ref{ham}) is completely integrable
\cite{C,M,OP} and the complete set of integrals of motion can be
represented as  spectral invariants of the Lax operator. Namely,
define $N\times N$ Krichever's Lax operator \cite{K} \be
{\cal{L}}(u)=\sum_{i=1}^Np_iE_{ii}-ig\sum_{i\neq
j}\Phi(u,x_i-x_j)E_{ij}\label{lax} \ee with matrix $E_{ij}$
having the following nonzero entries
$(E_{ij})_{kl}=\d_{ik}\d_{jl}$ and $\Phi(u,x)$ defined by
(\ref{phi}). Then a decomposition of ${\mbox{\rm det}}(z\cdot{\bf
1}-{\cal{L}}(u))$ in $z$ \be {\mbox{\rm det}}(z\cdot{\bf
1}-{\cal{L}}(u))= \sum_{i=0}^{N}(-1)^{i}z^{N-i}t_i(u) \ee
generates a set of commuting hamiltonians $H_i$, $i=1,\ldots,N$
with respect to the Poisson bracket \be \{H_i,H_j\}=0,\quad
i,j=1,\ldots,N. \ee

\section{The separation of variables}

In this section we briefly remind the results from the Sections 3
and 6 of \cite{KNS} (see also \cite{S}).

We are looking for a canonical transformation $K$ which maps
$(q,p)\mapsto(u,v)$, $H_i(x,p)\mapsto H_i(u,v)$ such that there
exist $N$ relations \be\Phi_i(u_i,v_i;H_1,\ldots,H_N)=0,\quad
i=1,\ldots,N. \label{sep} \ee

The main problem is to construct a generating function
${\cal{F}}(u|q)$ which performs such a separation.

A Baker-Akhiezer function is defined as the eigenvector of the Lax
operator ${\cal{L}}(u)$
\be
{\cal{L}}(u)\Omega(u)=v(u)\Omega(u) \label{baker}
\ee
with a
normalization fixed by a linear condition
\be
\vec{\a}\cdot\Omega\equiv\sum_{i=1}^N\a_i(u)\Omega_i(u)=1.
\label{norm}
\ee

The separated variables $u_i$ are thought as {\it poles of the
properly normalized} Baker-Akhiezer function. Then the canonically
conjugated variables $v_i$ are the corresponding eigenvalues of
${\cal{L}}(u_i)$ and satisfy separation equations (\ref{sep})
\be
\Phi_i\equiv\mbox{\rm det}(v_i\cdot{\bf 1}-{\cal{L}}(u_i))=0 \ee

From (\ref{baker}-\ref{norm}) it follows that
\be
\Omega(u)=\left(
\begin{array}{c}\vec{\a} \\
\vec{\a}{\cal{L}}(u)\\ \vdots\\
\vec{\a}{\cal{L}}^{n-1}(u)\end{array}\right)^{-1}\cdot \left(
\begin{array}{c}1 \\
v\\ \vdots\\ v^{n-1}\end{array}\right) \ee

Define the function
\be
\mathfrak{B}(u)=\mbox{\rm det}\left(
\begin{array}{c}\vec{\a} \\
\vec{\a}{\cal{L}}(u)\\ \vdots\\
\vec{\a}{\cal{L}}^{n-1}(u)\end{array}\right). \ee

Then the poles $u_i$ (or separated variables) of the
Baker-Akhiezer function are defined from the condition
$\mathfrak{B}(u_j)=0$.

It has been shown in \cite{KNS} that the simplest normalization
condition $\vec{\a}(u)=(0,0,\ldots,0,1)$ works for the
Calogero-Moser system. With such a normalization the expression
for $\mathfrak{B}(u)$ takes the form
\be
\mathfrak{B}(u)=\mbox{\rm det}\left(
\begin{array}{ccc}0 &\cdots & 1 \\{\cal{L}}_{n1} &\cdots&
{\cal{L}}_{nn} \\ \vdots &\ddots&\vdots \\ ({\cal{L}}^{n-1})_{n1}
&\cdots& ({\cal{L}}^{n-1})_{nn}
\end{array}\right). \label{bdet} \ee

Given the poles $u_i$ the conjugate variables $v_i$ can be defined
from the equation
\be
({\cal{L}}(u_i)-v_i)^\wedge_{nk}=0,\quad k=1,\ldots,n \label{vexp}
\ee and the wedge denotes the adjoint matrix.

It was shown in \cite{KNS} that in the primitive domain
${\mathfrak{D}}$ the function
$B(u)$ has exactly $N-1$ zeros $u_i$ and $N-1$ pair $(u_i,v_i)$
together with the variables $(Q,P)$, describing the motion of the
centre-of-mass,
\be
X=q_N,\quad P=\sum_{i=1}^N p_i \ee
give the complete canonical set
of new variables.

First let us examine  the $A_2$ case (see \cite{KNS}). We shall
introduce another sets of canonical variables. The first set
$(y_i;x_i,Q;P)$ simply describes a separation of the motion of the
centre-of-mass
\be
\begin{array}{lll}
x_1=q_1-q_3,& x_2=q_2-q_3,& Q=q_3 \\
y_1=p_1,&y_2=p_2,&P=p_1+p_2+p_3.
\end{array}
\ee

Then we introduce two sets of canonical variables in the reduced
phase space (with eliminated canonical variables $(Q,P)$)
\be
\begin{array}{llll}
x_+=x_1+x_2,& x_-=x_1-x_2,&
y_+={1\over2}(y_1+y_2),&y_-={1\over2}(y_1-y_2)\\ \phantom{-}&
\phantom{-}& \phantom{-}& \phantom{-}\\ u_+=u_1+u_2,&
u_-=u_1-u_2,& v_+={1\over2}(v_1+v_2),&v_-={1\over2}(v_1-v_2)
\end{array}\label{var}
\ee

The generating function $\mathcal{F}$ of the separating
transformation can be written as $\mathcal{F}(v_+,u_-;x_+,x_-)$ or
$\mathcal{F}(v_+,u_-;x_1,x_2)$. We prefer to use the second form
which is more convenient for a generalization to the $A_3$ case.
This function performs the canonical transformation from
$(x_{1,2},y_{1,2})$ to $(u_\pm,v_\pm)$ such that \be
u_1+u_2=x_1+x_2 \quad{ \mbox{\rm mod} \>\>\Gamma}\label{cons}\ee
and
\be
{\p\mf\over\p x_1}=y_1,\quad{\p\mf\over\p x_2}=y_2,\quad
{\p\mf\over\p v_+}=u_+,\quad{\p\mf\over\p
u_-}=-v_-.\label{red2}\ee

The next trivial observation is important for a generalization to
the $A_3$ case: The function $\mf(v_+,u_-;x_1,x_2)$ allows the
following decomposition
\be
\mf(v_+,u_-;x_1,x_2)=v_+x_++ig\log{\s (x_1)\s (x_2)\over \s (u_1)
\s (u_2)\s(x_1-x_2)}+\ov \mf(u_-,x_-). \label{dec} \ee

Here we imply that all variables in the RHS of (\ref{dec}) have to
be expressed in terms of $(v_+,u_-;x_1,x_2)$ using
(\ref{var}-\ref{cons}). Note that $\ov\mf$ depends only on
pairwise differences of $x_i$, $u_i$. Then we have
\be
y_{1,2}=v_++ig[\zeta(x_{1,2})\mp\zeta(x_1-x_2)-{1\over2}(\zeta(u_1)+
\zeta(u_2))]+\ov y_{1,2},\>\>\> \ov y_{1,2}={\ds\p \ov\mf\over\p
x_{1,2}}. \label{shiftp}\ee

Evaluating the determinant in (\ref{bdet}) and using (\ref{vexp})
we obtain that the condition $\mathfrak{B}(u)=0$ is equivalent to
\be
v_{1,2}=A_1(u_{1,2})=A_2(u_{1,2}) \label{surf} \ee with
\be
A_i(u)=y_i+ig[\zeta(u)-\zeta(x_i)+\zeta(x_i-x_{3-i})-\zeta(u-x_{3-i})].
\ee Using (\ref{shiftp}) we can rewrite (\ref{surf}) as follows
\be
\ov y_1-\ov y_2=2{\p\over\p{x_-}}\ov
\mf(u_-,x_-)=ig[\zeta(u-x_2)-\zeta(u-x_1)],\quad u=u_{1,2}
\label{red1}\ee

It is a simple calculation to check that a solution of
(\ref{red1}) which is compatible with
(\ref{cons}-\ref{red2},\ref{shiftp}-\ref{surf}) has the following
form
\be
\ov
\mf(u_-,x_-)=ig\log\s\bigl({x_-+u_-\over2}\bigr)\s\bigl({x_--u_-\over2}\bigr).
\ee We see that the partial differential equation (\ref{red1}) for
$\ov\mf$ involves a reduced number of variables and  looks rather
simpler than the equation (\ref{surf}) for $\mf$. Our purpose is
to obtain analogs of (\ref{red1}) for the $A_3$ case and try to
solve them.

Again we start with a set of canonical variables $(y_i;x_i,Q;P)$,
$i=1,2,3$
\be
\begin{array}{llll}
x_1=q_1-q_4,& x_2=q_2-q_4,& x_3=q_3-q_4,& Q=q_4 \\
y_1=p_1,&y_2=p_2,&y_3=p_3,& P=p_1+p_2+p_3+p_4\label{var3}
\end{array}
\ee and introduce in the reduced phase space canonical variables
\be
{\ds\begin{array}{lll}\ds x_+={{x_1+x_2+x_3}\over3},&\ds x'=
{2x_1-x_2-x_3\over3}
,&
\ds x''={2x_2-x_1-x_3\over3}\\ \phantom{.}&\phantom{.}&\phantom{.}\\\ds
y_+=y_1+y_2+y_3,&\ds y'=y_1-y_3,&\ds
y''=y_2-y_3
\end{array}}\label{var3a}\ee
and similarly a set of separated variables
\be
{\ds\begin{array}{lll}\ds u_+={{u_1+u_2+u_3}\over3},&
\ds u'={{2u_1-u_2-u_3}\over3},&
\ds u''={{2u_2-u_1-u_3}\over3}\\ \phantom{.}&\phantom{.}&\phantom{.}\\\ds
v_+=v_1+v_2+v_3,&\ds v'=v_1-v_3,&\ds
v''=v_2-v_3.\label{var3b}
\end{array}}\ee

The generating function $\mf(v_+,u',u'';x_1,x_2,x_3)$
performs the canonical transformation
from $(x_{1,2,3},y_{1,2,3})$ to $(u_+,u',u'';v_+,v',v'')$ such
that
\be
u_1+u_2+u_3=x_1+x_2+x_3\quad{ \mbox{\rm mod} \>\>\Gamma}
\ee
and
\be
{\p\mf\over\p v_+}=u_+,\quad{\p\mf\over\p
u'}=-v',\quad{\p\mf\over\p
u''}=-v'',\quad{\p\mf\over\p x_i}=y_i,\quad i=1,2,3.\label{red3}\ee

We introduce the ``reduced'' generating function ${\ov \mathcal F}$
by the formula
\be
\mf=v_+x_++ig\log{\ds\prod_{i=1}^3\s (x_i)\over \ds\prod_{i=1}^3\s
(u_i) \prod_{i<j}\s(x_i-x_j)}+ig\ov \mf. \label{red4} \ee

Zeros of the determinant (\ref{bdet})
define the separated variables
$u_i$, $i=1,2,3$. Then the conjugated variables $v_i$ are simply
rational functions of matrix elements of the Lax operator evaluated
at $u_i$ and can be found from (\ref{vexp}).
We want to find a convenient expression for this determinant.
This calculation is quite tedious and involves
complicated elliptic identities between Weierstrass functions.
The easiest way to calculate $\mathfrak{B}(u)$
is to check compatibility conditions
for $v_i$ coming from (\ref{vexp}) (see formulas (\ref{red45})
below). Here we shall only give the final result.

Using (\ref{red4}) let us make a change of variables
\be
y_{i}={1\over3}v_++ig[\zeta(x_i)-\zeta(x_i-x_j)-\zeta(x_i-x_k)-{1\over3}
\sum_{l=1}^3\zeta(u_l)]+ig\ov y_i, \label{red5} \ee where
$\{i,j,k\}$ is a permutation of $\{1,2,3\}$, ${\ds\ov
y_i={\p\over\p
  x_i}\ov\mf}$.

Then the determinant in (\ref{bdet}) can be written as \be
\mathfrak{B}(u)= {\ds ig^3\prod_{i=1}^3\Phi(u,-x_i)}\mathfrak
B(r_1,r_2|\vec x,u)\ee with \beq&\mathfrak B(r_1,r_2|\vec
x,u)=\Bigl\{\wt r_1\wt r_2(\wt r_1-\wt r_2)+&\nonumber\\&+2\wt
r_1\wt r_2[\zeta(x_1-u)-\zeta(x_2-u)-\zeta(x_1-x_2)]+&\nonumber\\
&+\wt
r_1^2[\zeta(x_1-x_2)+\zeta(x_2-u)-\zeta(x_1-x_3)-\zeta(x_3-u)]+
&\nonumber\\ &+\wt
r_2^2[\zeta(x_1-x_2)-\zeta(x_1-u)+\zeta(x_2-x_3)+\zeta(x_3-u)]\Bigr\},
&\label{red6} \eeq where $\vec{x}\equiv\{x_1,x_2,x_3\}$ and \beq
&\wt r_{1,2}=r_{1,2}+
2[\zeta(x_3-u)-\zeta(x_{1,2}-u)],&\nonumber\\ & {\ds r_{1,2}= \ov
y_{1,2}-\ov y_3=\Bigl\{{\p\over\p x'},{\p\over\p
  x''}\Bigr\}\ov\mf.}&
\label{red7}
\eeq

From (\ref{red6}) we can see that this equation depends only on
pairwise differences of $x_i$ and $u$ as in (\ref{red1}).
Therefore, the reduced generating function $\ov\mf$ depends
effectively on 4 independent variables (say, $x_i-u_1$, $i=1,2,3$
and $u_2-u_1$). However, despite the fact that a big
simplification happened we still have a very complicated nonlinear
partial differential equation (\ref{red6}) with elliptic
coefficients which is difficult to solve.

In the next sections we will show that a natural parameterization
of the equation (\ref{red6}) comes from the quantum case.

\section{The quantum $A_3$ Calogero-Moser system}

For the classical $A_3$ Calogero-Moser system
 we have four commuting hamiltonians
\cite{OP}
\beq
&{\ds H_1=\sum_{i=1}^4p_i,\quad
\ds H_2=\sum_{i<j}p_ip_j-g^2\sum_{i<j}\wp(q_i-q_j),}&\nonumber\\
&{\ds
  H_3=\sum_{i<j<k}p_ip_jp_k-g^2\sum_{i<j}\wp(q_i-q_j)(p_k+p_l),\quad
i<j\neq k<l}\\
&{\ds
  H_4=p_1p_2p_3p_4-g^2\sum_{i<j}\wp(q_i-q_j)[p_kp_l-{1\over2}
g^2\wp(q_k-q_l)],
\>\> i<j\neq k<l}.\label{red8}\nonumber
\eeq
They come from the spectral invariants of the Lax operator (\ref{lax})
\be
{\mbox{\rm det}}(z\cdot{\bf 1}-{\cal{L}}(u))=z^4-z^3t_1(u)+z^2t_2(u)-
zt_3(u)+t_4(u),\label{red9}
\ee
\beq
&t_1(u)=H_1,&\nonumber\\
&t_2(u)=H_2+6g^2\wp(u),&\nonumber\\
&t_3(u)=H_3+3g^2\wp(u)H_1-4ig^3\wp'(u),&\nonumber\\
&t_4(u)=H_4-ig^3\wp'(u)H_1+g^2\wp(u)H_2+g^4[3\wp^2(u)-\wp''(u)].&
\label{red10}
\eeq
The separated variables $(v_j,u_j)$ satisfy the relations
\be
{\mbox{\rm det}}(v_j\cdot{\bf 1}-{\cal{L}}(u_j))=
v_j^4-v_j^3t_1(u_j)+v_j^2t_2(u_j)-
v_jt_3(u_j)+t_4(u_j)=0.\label{red11}
\ee

Now let us consider the quantum case. We replace $p_i$ by differentiations
$p_j\to-i\p_{q_j}$ and instead of hamiltonians (\ref{red8})
we have four commuting differential operators
\beq
&{\ds H_1=-i\sum_{j=1}^4\p_{q_j},\quad
\ds H_2=-\sum_{j<k}\p_{q_j}\p_{q_k}-
g(g-1)\sum_{j<k}\wp(q_j-q_k),}&\nonumber\\
&{\ds
  H_3=i\sum_{j<k<l}\p_{q_j}\p_{q_k}\p_{q_l}+i
g(g-1)\sum_{j<k}\wp(q_j-q_k)(\p_{q_l}+\p_{q_m}),}\label{red12}\\
&{\ds
  H_4=\p_{q_1}\p_{q_2}\p_{q_3}\p_{q_4}
+g(g-1)\sum_{j<k}\wp(q_j-q_k)[\p_{q_l}\p_{q_m}+
{g(g-1)\over2}\wp(q_l-q_m)],}
\nonumber
\eeq
where $j<k\neq l<m$.

As it explained in \cite{S} for the $A_2$ case
(see also \cite{KS} for a trigonometric
case) the idea is to construct the linear operator $K$ which
intertwines $\{q_i\}$ and $\{u_i;Q\}$ representations.

Namely, we are looking for  the kernel
$K(\vec{u},Q;\vec{q})$, $\vec{u}=\{u_1,u_2,u_3\}$,
$\vec{q}=\{q_1,q_2,q_3,q_4\}$ of the operator $K$
such that
\be
K(\vec{u},Q;\vec{q})=\d(Q-q_4)
\wt K(\vec{u};\vec{x}),\label{red13}
\ee
where the variables $u_i,x_i$ are defined by (\ref{var3}-\ref{var3b}).
The spectral determinant (\ref{red11}) is replaced by
the following differential equation for the kernel $K$
\beq
&[\p_{u_j}^4-iH^*_1\p_{u_j}^3-[H^*_2+
6g(g-1)\wp({u_j})]\p_{u_j}^2+&\nonumber
\\ &+[iH^*_3+3ig(g-1)H^*_1\wp({u_j})+
4g(g-1)(g-2)\wp'({u_j})]\p_{u_j}+&\nonumber
\\ &+H^*_4+g(g-1)H^*_2\wp({u_j})-
ig(g-1)(g-2)H^*_1\wp'({u_j})+&\nonumber \\
&+3g^2(g-1)^2\wp^2({u_j})- g(g-1)(g^2-3g+3)\wp''({u_j})]K=0&
\label{red14}\eeq
where $H_i^*$ is the Lagrange adjoint of $H_i$
\be
\int\phi(\vec{q})(H\psi)(\vec{q})d\vec{q}=
\int(H^*\phi)(\vec{q})\psi(\vec{q})d\vec{q}\label{red15}
\ee
and the condition $P=-i\p_Q$ is replaced by
\be
[-i\p_Q-H_1^*]K=0\label{red16}
\ee
which is trivially satisfied because of (\ref{red13}).

One of possible ways to fix coefficients in (\ref{red14}) is to
look at two different limits: the classical one (when $g\to\infty$
and we should have  (\ref{red11})) and the trigonometric limit
$\wp(x)\to\csc(x)^2$, $\zeta(x)\to\cot(x)$, where
the version of (\ref{red11}) has been conjectured in
\cite{KS}. These two limits fix coefficients in (\ref{red14})
uniquely.

Let us assume that $\Psi(\vec{q})$ is an eigenfunction of $H_i$,
$i=1,2,3,4$ and consider the integral transform
\be
\wt\Psi(\vec{u},Q)=\int d{\vec{q}}\>K(\vec{u},Q;\vec{q})
\Psi(\vec{q})\label{red17} \ee Now we demand that the function
$\wt\Psi(\vec{u},Q)$ should satisfy the separated equations
\be
[-i\p_Q-h_1]\wt\Psi(\vec{u},Q)=0,\quad \mathfrak{D}_{u_j}
\wt\Psi(\vec{u},Q)=0,\label{red18}
\ee
where
\beq
&\mathfrak{D}_y=\p_y^4-ih_1\p_y^3-[h_2+6g(g-1)\wp(y)]\p_y^2+&\nonumber
\\ &+[ih_3+3ig(g-1)h_1\wp(y)+4g(g-1)(g-2)\wp'(y)]\p_y+&\nonumber
\\ &+h_4+g(g-1)h_2\wp(y)-ig(g-1)(g-2)h_1\wp'(y)+&\nonumber \\
&+3g^2(g-1)^2\wp^2(y)- g(g-1)(g^2-3g+3)\wp''(y),& \label{red19}\eeq
and $h_i$ are the eigenvalues of $H_i$ corresponding to the
eigenfunction $\Psi(\vec{q})$.

We are not going to discuss in this paper the question
of correct boundary conditions for the operator $K$ and differential
equation (\ref{red19}). We only make an assumption that
the boundary can be chosen in such a way that it does not
contribute to the result while integrating by parts
using (\ref{red14}-\ref{red15}). Unlike to the $A_2$ case
a correct choice of boundary conditions for (\ref{red17})
appears to be quite a complicated problem even in the
trigonometric limit and we will address this problem in a separate paper.

Our purpose is to solve exactly the differential equation
(\ref{red14}) for the kernel $K$. Substituting adjoints of
hamiltonians $H_i$ (\ref{red12}) into (\ref{red14}), using a
factorization (\ref{red13}) of the kernel $K$ and the change of
variables (\ref{var3}) we come to the following equation \be
\mathfrak{D}^{(1)}_{u_j}({u_j};\vec{x})\wt
K(\vec{u};\vec{x})\>\p_Q+
\mathfrak{D}^{(0)}_{u_j}({u_j};\vec{x})\wt K(\vec{u};\vec{x})=0
\label{red20} \ee where $\mathfrak{D}^{(1)}_{u_j}({u_j};\vec{x})$
and $\mathfrak{D}^{(0)}_{u_j}({u_j};\vec{x})$ are the 3-rd and
4-th order differential operators in $u_j$, respectively. The
kernel $\wt K(\vec{u}; \vec{x})$ should solve both of them.

Now guided by the $A_3$ classical case let us make a
substitution
\be
\wt K(\vec{u};\vec{x})=\Biggl\{{\ds\prod_{i=1}^3{\s(x_i)\over\s(u_i)}
\over\ds
\prod_{i<j}\s(x_i-x_j)}\Biggr\}^{g-1}\wt L(\vec{u};\vec{x})\label{red21}
\ee
and assume that the reduced kernel $\wt L(\vec{u};\vec{x})$ has
the following invariance
\begin{conj}\label{invar}
\be
\wt L(u_1+\l,u_2+\l,u_3+\l;x_1+\l,x_2+\l,x_3+\l)=\wt L(\vec{u};\vec{x}),
\quad \forall\l\in\mathfrak{D}.\label{red22}
\ee
\end{conj}
It appears that both equations in (\ref{red20}) are compatible
with (\ref{red22}) provided that $\wt L(\vec{u};\vec{x})$
satisfies the following system of linear partial differential
equations with elliptic coefficients \beq &\Bigl\{(g-1)^2
[\wp(x_\a-x_\b)-\wp(x_\g-u_i)-\zeta^2(x_\a-x_\b)+\zeta^2(x_\g-u_i)]+
&\nonumber\\ &+(g-1)[\zeta(x_\a-x_\b)(\p_{x_\a}-\p_{x_\b})+
\zeta(x_\g-u_i)(2\p_{u_i}+\p_{x_\a}+\p_{x_\b})]+&\nonumber\\
&+(\p_{u_i}+\p_{x_\a})(\p_{u_i}+\p_{x_\b}) \Bigr\} \wt L(\vec u;\vec
x)=0,\quad i,\>\a<\b=1,2,3.&\label{red23} \eeq

Again the differential operator in (\ref{red23}) of the second
order is considerably simpler than differential operators
in (\ref{red20}). The statement that the kernel $\wt K$ (\ref{red21})
with $\wt L$ satisfying (\ref{red23}) will solve (\ref{red20})
can be proved by direct calculations (very lengthy).
In fact, using the {\bf Conjecture \ref{invar}}
the equations (\ref{red23}) can be obtained
only from the equation
$\mathfrak{D}^{(1)}_{u_j}({u_j};\vec{x})\wt K(\vec{u};\vec{x})=0$.
Then the second equation
$\mathfrak{D}^{(0)}_{u_j}({u_j};\vec{x})\wt K(\vec{u};\vec{x})=0$
is valid automatically.

We strongly believe that $\wt K$ with the factorization (\ref{red21})
and  $\wt L$ satisfying (\ref{red23}) is the only sensible
solution to (\ref{red20}). However, it would be very interesting
to find other solutions to (\ref{red20}) which are not of the
form (\ref{red21}).

Now we will solve the system (\ref{red23}) for the kernel $\wt
L(\vec{u};\vec{x})$.

\begin{thm}\label{reduct}
The kernel $\wt L(\vec{u};\vec{x})$ admits further factorization
\be
\wt L(\vec{u};\vec{x})=\d(u_+-x_+)L(\vec{t},s),\label{red24}
\ee
\be
\vec{t}=\{t_1,t_2,t_3\},\quad t_i=x_i-u_1,\quad
s=u''-u'=u_2-u_1\label{red25}
\ee
 with $u_+,u',u''$ defined in (\ref{var3b}).
\end{thm}
{\bf Proof}:

Consider (\ref{red23}) for $i=1,2,3$
and rewrite it terms of variables $t_i=x_i-u_1$, $s=u_2-u_1$, $u_1$  and
$\Delta=u_+-x_+$. Using the {\bf Conjecture \ref{invar}}
and comparing mixed derivatives
of $\wt L$ one can show that (\ref{red23}) is compatible only if
\be
\wt L(u_1,s;t_1,t_2,t_3;\Delta)\sim\d(\Delta)
\ee.
\hfill $\blacksquare$

Now introduce differential operators
\beq
&\mathfrak{D}_{\a\b}\equiv\p_{t_\a}\p_{t_\b}+(g-1)
[\zeta(t_\a-t_\b)(\p_{t_\a}-\p_{t_\b})+&\nonumber\\&
+\zeta(s-t_\a-t_\b)(\p_{t_\a}+\p_{t_\b})]+(g-1)^2\times&\nonumber\\
&\times[\wp(t_\a-t_\b)-\wp(s-t_\a-t_\b)-\zeta^2(t_\a-t_\b)+
\zeta^2(s-t_\a-t_\b)], &\label{red27} \eeq \beq
&\mathfrak{D'}_{\a\b}\equiv(\p_{t_\a}+\p_s)(\p_{t_\b}+\p_s)+&\nonumber\\
&+(g-1)[\zeta(t_\a-t_\b)(\p_{t_\a}-\p_{t_\b})+
\zeta(t_\g-s)(\p_{t_\a}+\p_{t_\b}+2\p_s)+&\nonumber\\
&+(g-1)^2[\wp(t_\a-t_\b)-\wp(t_\g-s)-\zeta^2(t_\a-t_\b)+\zeta^2(t_\g-s)],
\label{red28} & \eeq
\beq
&\mathfrak{D''}_{\a\b}\equiv(\p_{t_\a}+\p_{t_\g}+\p_s)
(\p_{t_\b}+\p_{t_\g}+\p_s)+&\nonumber\\
&+(g-1)[\zeta(t_\a-t_\b)(\p_{t_\a}-\p_{t_{\b}})-
\zeta(t_\g)(\p_{t_\a}+\p_{t_\b}+2\p_{t_\g}+2\p_s)]+&\nonumber\\
&+(g-1)^2[\wp(t_\a-t_\b)-\wp(t_\g)-\zeta^2(t_\a-t_\b)+\zeta^2(t_\g)],
\label{red29} & \eeq where $\a,\b,\g$ is a permutation of
$\{1,2,3\}$.
Then the system (\ref{red23}) for the kernel
$\wt L(\vec{u};\vec{x})$ is equivalent to the following
system of equations for $L(\vec{t},s)$
\be
\mathfrak{D}_{\a\b}L(\vec{t},s)=0,\quad
\mathfrak{D'}_{\a\b}L(\vec{t},s)=0,\quad
\mathfrak{D''}_{\a\b}L(\vec{t},s)=0.\label{red30}
\ee
The following theorem is an elliptic generalization of the result
given in \cite{VM}
\begin{thm}\label{solution}
A solution for the system (\ref{red30}) is given by the following
expression
\be
{L}(\vec{t},s)={\ds\oint_{\mathcal{C}}}{dz}\> \tau(\vec{t},s|z)
\label{red31} \ee where
\be
\tau(\vec{t},s|z)=\varkappa(\vec{t},s|z)^{g-1}, \label{red32}
\ee
\be
\varkappa(\vec{t},s|z)={\s(z)\s(z+s)\over\s^2(2z+s)}\prod_{i=1}^3
\s(z+t_i)\s (z+s-t_i) \label{red33} \ee and the contour
$\mathcal{C}$ is closed on the Riemann surface of the integrand.
\end{thm}
{\bf Proof:}
The proof of the theorem is based on three elliptic identities:
\be
\mathfrak{D}_{\a\b}\bigl[\tau(\vec t,s|z)\bigr]=0,\label{red34}
\ee
\be
\mathfrak{D'}_{\a\b}\bigl[\tau(\vec t,s|z)\bigr]=(g-1){\p\over\p
z} \Biggl[ {\s(z)\s(z+t_\g)\s(2z+2s-t_\g)\tau(\vec
t,s|z)\over\s(z+s-t_\g)\s(z+s)\s(2z+s)\s(t_\g-s)}
\Biggr]\label{red35} \ee and
\be
\mathfrak{D''}_{\a\b}\bigl[\tau(\vec t,s|z)\bigr]=(g-1){\p\over\p
z} \Biggl[ {\s(z)\s(t_\g-z-s)\s(2z+s+t_\g)\tau(\vec
t,s|z)\over\s(z+s)\s(2z+s)\s(t_\g)\s(z+t_\g)} \Biggr].
\label{red36}\ee

Formulas (\ref{red34}-\ref{red36}) can be proved either by using
(\ref{add1}-\ref{addth}) or checking that a difference of LHS
and RHS are elliptic functions with no poles.

These identities show that under the action of $\mathcal{D}_{\a\b}$,
$\mathcal{D'}_{\a\b}$, $\mathcal{D''}_{\a\b}$ the integral in (\ref{red31})
becomes a total derivative of the function with the same singularities
as the function $\tau(\vec{t},s|z)$ in (\ref{red32}).
\begin{flushright}
$\blacksquare$
\end{flushright}
A natural question arizes: do (\ref{red31}-\ref{red33}) describe
a general solution to the system (\ref{red30}) ? In the
trigonometric limit the answer is positive and changing the
contour $\mathcal{C}$ we can produce the whole basis of linearly
independent solutions \cite{VM}.
It is likely that this statement can be
generalized to the elliptic case as well.

In fact, all we proved that if we choose integration contours to
be some curves in (\ref{red31}) and (\ref{red17}) closed on the
Riemann surface of the integrands, then the equations
(\ref{red18}) should be valid. Of course, it does not guarantee
that the function (\ref{red17}) will split into the product of
functions depending on $Q$ and $u_i$ separately. However,
the asymptotics of the integral (\ref{red31}) in
the classical limit $g\to\infty$ provides a natural
parameterization for (\ref{red6}).

So let us consider the limit $g\to\infty$. Then we have to
calculate the asymptotic behaviour of the reduced kernel
$L(\vec{t},s)$ when $g\to\infty$. It is clear that, in general,
this asymptotics is a multi-valued function of $(t_1,t_2,t_3,s)$.
Due to a special form (\ref{red32}) of $\tau(\vec{t},s|z)$ we can
use the steepest descent method to obtain that \be
{L}(\vec{t},s)|_{g\to\infty} \backsimeq
\exp(g\log\mathfrak{F}(\vec{t},s))\label{red37}\ee where \be
\mathfrak{F}(\vec{t},s)=\k(\vec{t},s|z^*)\label{red38} \ee
with
\be {\ds\p\over\ds\p
  z}\k(\vec{t},s|z)|_{z=z^*}=0.\label{red39} \ee

We can rewrite the equation (\ref{red39}) for $z^*$ as
\be
\zeta(z^*)+\zeta(z^*+s)+\sum_{i=1}^3[\zeta(t_i+z^*)+\zeta(s-t_i+z^*)]=
4\zeta(s+2z^*) \label{red40} \ee

This equation defines the stationary phase point $z^*$ at which
the function (\ref{red38}) has to be evaluated.

\section{The $A_3$ generating function}

We shall start with the following { Lemma} which provides the main
technical tool for the constructing of the $A_3$ generating
function

\begin{lem}\label{para}

The function $\k(\vec{t},s|z)$ satisfies the following partial
differential equation with elliptic coefficients \beq &\mathfrak
B({\ds\p\over\ds \p x'}\log\k(\vec t,s|z),{\ds\p\over \ds \p
x''}\log\k(\vec t,s|z)|\vec t,v)=&\nonumber\\ &={\ds{\p\over\p
z}\k(\vec t,s|z)}
{\ds\s(\sum_{i=1}^3t_i-s-v)\s(z)^2\s(z+s)^2\prod_{i<j}\s(t_i-t_j)\over
\ds\k(\vec t,s|z)^2\s(2z+s)\prod_{i=1}^3\s(t_i-v)},
&\label{red41}\eeq where $v=\{0,s,t_1+t_2+t_3-s\}$, $t_i=x_i-u_1$,
$s=u_2-u_1$ and the function ${\mathfrak B(r_1,r_2;\vec x,u)}$
defined by relations (\ref{red6}-\ref{red7}).
\end{lem}

The proof of the Lemma is straightforward, but technically
complicated. It is instructive to start with the case
$v=t_1+t_2+t_3-s$, when the RHS in (\ref{red41}) is zero. Then the
LHS is some combination of Weierstrass $\zeta$ functions which is
zero. The first two cases $v=\{0,s\}$ are ones of the most
complicated elliptic identities in this paper.
They can be proved in several steps using identities similar to
(\ref{red45}) and (\ref{red47}-\ref{red49}) (see below).

Now we are ready to formulate the main result of this paper
\begin{thm}\label{main}
The $A_3$ generating function $\mf(v_+,u',u'';x_1,x_2,x_3)$
performing the canonical transformation from
$(x_{1,2,3},y_{1,2,3})$ to $(u_+,u',u'';v_+,v',v'')$ is given by
\be
\mf=v_+x_++ig\log{\ds\prod_{i=1}^3\s (x_i)\over \ds\prod_{i=1}^3\s
(u_i) \prod_{i<j}\s(x_i-x_j)}+ig\ov \mf. \label{red42} \ee
 with $\ov \mathcal{F}(\vec{t},s)$
\be
\ov \mathcal{F}(\vec t,s)=\log\mathfrak F(\vec t,s),\label{red43}
\ee where $\mathfrak F(\vec t,s)$ is defined by
(\ref{red38}-\ref{red40}), the variables $\vec t,s$ by
(\ref{red25}) and all variables $u_i,u_+,u',u''$,
$v_i,v_+,v',v''$, $x_i$, $y_i$ by (\ref{var3}-\ref{var3b}).
\end{thm}
{\bf Proof:} The proof proceeds in two steps. First of all we have
to check that with   the generating function (\ref{red42}) the
equation $\mathfrak B(u)=0$  has three roots $u_1,u_2,u_3$ in the
primitive domain $\mathfrak D$. Now using {\bf
Lemma \ref{para}} and definitions (\ref{var3}-\ref{red7}) it is
easy to see that three roots $u_1,u_2,u_3$ correspond exactly to
the cases $v=\{0,s,t_1+t_2+t_3-s\}$ of {\bf Lemma \ref{para}}. So
choosing $z^*$ such that ${\ds\p\over\ds\p
  z}\k(\vec{t},s|z)|_{z=z^*}=0$
we obtain the solution to $\mathfrak B(u)=0$.

The next step is to show that the conjugated variables $v_1,v_2,v_3$
(or $v_+,v',v''$) defined by (\ref{vexp}) are compatible with
(\ref{red42}).

Let us denote as $v_1^*,v_2^*,v_3^*$ the conjugated variables
obtained from the equations \be {\p\over\p u'}\mathcal F =
-v_1^*+v_3^*,\quad {\p\over\p u''}\mathcal F = -v_2^*+v_3^*
\label{red43a} \ee and \be y_+=\sum_{i=1}^3v_i^*
+ig(\sum_{i=1}^3[\zeta(x_i)-\zeta(u_i)]+ \sum_{i=1}^3\ov
y_i),\label{red43b} \ee
where we simply used (\ref{red5}) and \beq
&{\ds\ov y_i={\p\over\p x_i}{\ov \mathcal F}=
\zeta(z+x_i-u_1)-\zeta(z+u_2-x_i)-}&\nonumber\\
&{\ds-{1\over3}\sum_{i=1}^3[
\zeta(z+x_i-u_1)-\zeta(z+u_2-x_i)]}&\label{red43c}
\eeq
from the formula for ${\ov\mathcal F}=
\log\k(\vec t,s|z)$. Note that
${\ds\sum_{i=1}^3\ov y_i=0}$
simply reflects the fact that ${\ov\mathcal F}$
depends only on four variables $u',u'',x',x''$.

Substituting (\ref{red42}) into (\ref{red43a}) and using
(\ref{red43b}-\ref{red43c})
we obtain the following expressions for  $v_i^*$
\beq
&{\ds v_1^*={1\over3}y_++ig\Bigl[\zeta(u_1)+\zeta(z+u_2-u_1)-
2\zeta(2z+u_2-u_1)+}&\nonumber\\
&{+\ds{1\over3}\sum_{i=1}^3[
\zeta(z+u_2-x_i)+2\zeta(z+x_i-u_1)-\zeta(x_i)]\Bigl]},&
\eeq
\beq
&{\ds v_2^*={1\over3}y_++ig\Bigl[\zeta(u_2)-\zeta(z+u_2-u_1)+
2\zeta(2z+u_2-u_1)+}&\nonumber\\
&{+\ds{1\over3}\sum_{i=1}^3[
-2\zeta(z+u_2-x_i)-\zeta(z+x_i-u_1)-\zeta(x_i)]\Bigl]},&
\eeq
\be
v_3^*={1\over3}y_++ig\Bigl[\zeta(u_3)+{1\over3}\sum_{i=1}^3[
\zeta(z+u_2-x_i)-\zeta(z+x_i-u_1)-\zeta(x_i)]\Bigl].
\ee

Now we check that these expressions are compatible with
(\ref{vexp}). The conditions (\ref{vexp}) were analyzed in
\cite{KNS} where many different expressions for $v_i$ were
obtained. All these expressions are equivalent provided that
$\mathfrak B(u_i)=0$. Let us introduce  matrices \cite{KNS} \be
{\cal L}^{(p)}(u)={\cal L}(u)[\mbox{\rm tr}{\cal L}^{(p-1)}(u)]-
(p-1){\cal L}^{(p-1)}(u){\cal L}(u),\quad {\cal L}^{(1)}(u)={\cal
L}(u). \ee Then from formula (3.23) of \cite{KNS} with $n=4$,
$i=1$, $j=3$, $k=\a$, $\a=1,2$ we have
\be
v_i={{\cal L}^{(1)}_{4\a}(u_i){\cal L}^{(3)}_{43}(u_i)- {\cal
L}_{43}^{(1)}(u_i){\cal L}_{4\a}^{(3)}(u_i)\over {{\cal
L}_{4\a}^{(1)}(u_i){\cal L}^{(2)}_{43}(u_i)- {\cal
L}_{43}^{(1)}(u_i){\cal L}_{4\a}^{(2)}(u_i)}}. \label{red44} \ee

Now using the  definition (\ref{lax}) of the Lax operator ${\cal
L}(u)$ and the addition theorem (\ref{addth}) one can rewrite
(\ref{red44}) in the following form \beq &v_i=y_\a+ig\Bigl[
\zeta(x_3-u_i)+\zeta(u_i)-\zeta(x_\a)+\zeta(x_\a-x_3)+&\nonumber\\
&+{\ds \wt r_{\a_{\phantom{i}}}(u_i)\over
\ds {\wt r_{3-\a}}^{\phantom{i}}(u_i)}
\bigl[\zeta(u_i-x_3)-\zeta(u_i-x_{3-\a})+\zeta(x_\a-x_{3-\a})-
\zeta(x_\a-x_3)\bigr] \Bigr],\quad\>\>&\label{red45} \eeq
\be
\wt r_\a(u_i)=\ov y_\a-\ov y_3+2\zeta(u_i-x_\a)-2\zeta(u_i-x_3),\quad
\a=1,2.\label{red46} \ee

It is easy to see that two expressions (\ref{red45}) for $\a=1,2$
are equivalent exactly when $\mathfrak B(u_i)=0$. Let us use
(\ref{red45}) with $\a=1$.
The following three elliptic identities can be proved using
(\ref{addth}). In fact, they are very useful
in the proof of {\bf Lemma  \ref{para}}.
We shall put them in the form convenient for  our purposes, namely,
\beq
&v_1-v_1^*=
ig{\ds1\over
\ds {\wt r_{2}}^{\phantom{i}}(u_1)}{\ds{\p\over\p z}\log\k(\vec t,s|z)}
\times&\nonumber\\
&\times
[\zeta(x_2-u_1)-\zeta(x_3-u_1)-\zeta(z+x_2-u_1)+\zeta(z+x_3-u_1)],&
\label{red47}
\eeq
\beq
&v_2-v_2^*=
ig{\ds1\over
\ds {\wt r_{2}}^{\phantom{i}}(u_2)}{\ds{\p\over\p z}\log\k(\vec t,s|z)}
\times&\nonumber\\
&\times
[\zeta(u_2-x_2)-\zeta(u_2-x_3)-\zeta(u_2-x_2+z)+\zeta(u_2-x_3+z)],&
\label{red48}
\eeq
\be
v_3-v^*_3 =0. \label{red49}
\ee
These elliptic
identities show that if we choose again $z$ to be  $z^*$ such that
${\ds{\p\over\p z}\log\k(\vec t,s|z)|_{z=z^*}}=0$, then
the generating function $\mathcal F(v_+,u',u'';x_1,x_2,x_3)$ satisfies
\be
d(\mathcal{F}-v_+u_+)=y_1dx_1+y_2dx_2+y_3dx_3-v_+du_+-v'du'-v''du''.
\ee
This identity proves that the transformation from
$(x_1,x_2,x_3,y_1,y_2,y_3)$ to $(u_+,u',u'';v_+,v',v'')$
is canonical \cite{A}.
\begin{flushright}
$\blacksquare$
\end{flushright}

\section{Conclusion}
In this paper we have constructed the generating function of the
canonical separating transform for the $A_3$ classical
Calogero-Moser system. This function appears to be multi-valued.
The approach we used originated from the quantum version of the
model. In fact, the purpose of this note was to show that the
conjectured quantum separating operator produces the correct
asymptotics in the classical limit. It adds us a self-confidence
that we obtained the  correct expression for the quantum kernel.
However, the problem of  correct boundary conditions  looks
complicated because of quite nontrivial monodromy properties of
this operator. We also think that it is straightforward to
generalize the results of this paper for the classical
Ruijsenaars system in line with \cite{KNS}. We hope to address
these problems in further publications.

Of course, a generalization of these results even for the
classical $A_n$ ($n>4$) case would be of a great interest.
The classical $A_3$ Calogero-Moser model appears to be the first case
when the generating function is a function
``living'' on some  complicated Riemann surface. However, we believe
that a consideration of the classical case can give a key
how to construct the $A_n$ quantum separating  operator.

\section{Acknowledgments}

I am grateful  to Vadim Kuznetsov and Jan DeGier for their
interest and very stimulating discussions and suggestions. Also I
would like to thank  Sergey Sergeev for discussions on the quantum
separation variables method. This research was supported by the
Australian Research Council.

\end{document}